Department of Science and Technology Studies

University College London

Science, Technology, and Society MSc Dissertation

# Corporate Disruption in the Science of Machine Learning

By: Sam Work

September 1, 2016

# Table of Contents

# Introduction

The infiltration of artificial intelligence (AI) into everyday life has been slowly and steadily increasing for over half a century. While the term often conjures fantastic rhetoric about the 'near-future' capabilities of robots and driverless cars, the more mundane aspects of the science continue to progress, improving online search results and identifying potential criminals (Allworth, 2015; Altman & Musk, 2015; Brustein, 2016). In doing so this area of research has attracted hundreds of millions of dollars, from early-stage investors wanting to fund the next big thing to large corporations looking to acquire their piece of the latest technology (Waters, 2015).

The sheer reach of the field in everyday life, its colourful billionaires, political clout and access to vast financial resources demand some scrutiny into the research assumed to be underpinning these advancements. Science and technology studies (STS) aims to open up such 'black boxes' which are often ascribed a set of assumptions and put aside. In doing so, it seeks to problematize the various explanatory narratives put forth by other disciplines—economics, political science, sociology, and often, itself—and greets anything presented in too linear a fashion with suspicion.

With that rather ambitious end in mind, this dissertation focuses on examining the changes occurring in machine learning research in the last decade, as experienced by the researchers currently working in it. This subfield of AI is at the current centre of attention for its contributions to the abovementioned developments. The empirical work of this dissertation is formed around 12 semi-structured interviews conducted with machine learning researchers working in academia and industry.

In the tradition of STS, this dissertation proposes theories and frameworks from a variety of disciplines in order to better understand and contextualize researchers' responses, while also critically assessing the validity of those theories and frameworks. Broadly it examines the impact of industry conducting science in the field has on research opportunities, practices, and academia itself. These are presented both on a micro and macro level, A brief history of the development of the field is presented both to contextualize interviewees' responses and to avoid the pitfalls of being "presentist" (Croissant & Smith-Doerr, 2008) by ignoring historical patterns and assessments.

## Literature Review

The current iteration of renewed interest in AI research, along with the high valuations of companies using machine learning and the enticing salaries commanded by computer scientists, has been well documented in the media (A.E.S., 2016; Bilton, 2014; "Don't be evil, genius," 2014, "Million-dollar babies," 2016, "Rise of the machines," 2015; Gibney, 2016; Kelly, 2014; Lohr, 2012; Markoff, 2016; Regalado, 2014; Waters, 2015). However, a survey of contemporary STS literature on the topic turns up surprisingly little. Classic STS philosophies of science (e.g., Popper, Kuhn, Merton) have been used to schematize the discipline (Dodig-Crnkovic, 2002). Collins (1995) provided an overview of how STS practitioners seek to understand science carried out via "intelligent machines". There have been previous laboratory studies, such as on expert systems in the 1980s (Forsythe, 1993). A survey of over 700 AI researchers globally looked at the effect increased interest in neural networks had on researchers' area of focus, in an attempt to tease the effects that 'hype' had on researcher motivations (Rappa & Debackcre, 1990). While STS theories have often been developed from and applied to the life sciences, particularly biotechnology (Fochler, 2016; Lacy, Glenna, Biscotti, Welsh, & Clancy, 2014; Moore, Kleinman, Hess, & Frickel, 2011; Vallas & Kleinman, 2007), fewer have engaged with contemporary computer science and AI research. Both fields have undergone periods of intense commercial investment and cycles of expectations and disappointment, but there are important differences, particularly when focusing on neoliberal effects, which highlight the need for further STS forays into the field.

A large amount of literature both within STS and other disciplines is concerned with the neoliberalization[1] of academia (Croissant & Smith-Doerr, 2008). The term's definition varies (Birch, 2016), but is commonly understood as "the promotion of market-based solutions to a broad range of issues" (Lave, Mirowski, & Randalls, 2010, p. 661). This literature takes as a common starting assumption that neoliberalism, induced through government policies (notably in the US, but also elsewhere), has infiltrated academia (Berman, 2014). Its beginnings are often traced back to the ascension of the military funding of science during World War II and the Cold War, followed by increased privatization and commercialization of 'knowledge' in the 1980s (Lave et al., 2010). Indications of this include an increased pressure to commercialize academic research via start-ups, patents, and other technology transfer processes. The "triple helix" model, which looks at the relationships between industry, academia and government, has been applied to analyzing innovation at companies in the field such as Google (Steiber & Alänge, 2013). Holloway (2015) considered the extent to which academic scientists exhibit agency in this environment. Moore et al. (2011) wrote about how prioritizing technology transfer from universities allowed "culture of commerce" to proliferate in academia. Suarez-Villa (2009) wrote that corporations had systematized experimentation to constantly produce new innovations in the service of profit and power, resulting in "technocapitalism".

---

[1] It has been argued this has been more a prevalent focus in the American stream of STS than the British for historic reasons (Mirowski & Sent, 2002).

Proponents of neoliberalism as a new ordering regime in science argue that its practice, beginning in the 1980s, is *exceptional* due to the resulting decreases in government funding for academic institutions, the isolation of research activities from teaching, and the subsequent intellectual property protection which thwarts the dissemination of research (Callon, 2002; Lave et al., 2010). However, there have also been sharp criticisms of this as reductivist and economic determinism. Different theories of the industrialization of research are criticized in STS for "treating knowledge itself as a black box easily handed off between university and industry scientists. The lack of epistemological sensibilities, such as how knowledge is constructed, is seen as a major weakness" (Croissant & Smith-Doerr, 2008, p. 703). Authors such as Shapin (2008) argue that science has never been free of financing and commercialization concerns.

Furthermore, neoliberal discussions often centre on commercialization activities of universities or public research which are not necessarily applicable to fields in computer science. For instance, while patents and licenses provide the focus for much of the literature on commercialization in the life sciences, they are not a particularly important mechanism for technology transfer in computer science (Perkmann et al., 2013).

There are other examinations of the industry/academia divide among researchers which focus on resource trading between the two in addition to commercial technology transfer, although this work is often under the supposition of neoliberalism. Slaughter, Campbell, Holleman and Morgan (2002) discussed the trade of graduate students by professors to industry as a means to forge ties and gain access, but in doing so relied heavily on the basic/applied research dichotomy. Lam (2011) explored researchers' motivations in the sciences, focusing on extrinsic validation through mechanisms like peer review, and intrinsic through discovery and money. Fochler (2016, p. 9) proposed the concept of "epistemic capitalism to denote the accumulation of capital through the act of doing research" which looks at how forms of worth (capital) is acquired in a research system. Of course, criticism of STS studies has been that *too much* weight is placed on researcher agency to the exclusion of other factors (for instance see Holloway, 2015). Other STS areas of inquiry, such as studies of hype and the dynamics of expectations, have also been adopted to aid with analysis in this dissertation (Borup, Brown, Konrad, & Van Lente, 2006). Additionally, classic STS texts (e.g., Polanyi, 1969) remain instructive and are subsequently referred to.

# Methodology

The novel research in this dissertation comes from narratives provided by machine learning researchers through semi-structured qualitative interviews. The goal here is to focus on the different ways researchers express their current situation in the field with relation to internal and external motivations. This approach has been used elsewhere (see Vallas & Kleinman, 2007), and aims to highlight narratives which "enable us to contextualize networks in the political economy" (Slaughter, 2001, p. 406). Rather than attempt to draw conclusions about the differences between industry and academic researchers across the field, the interviews are used here to highlight the diversity of individual experiences, while implying certain trends (Elliott, 2005).

Speaking with both researchers in industry and academia, including those who have traversed between the two and those who have just left one to begin in another, allows for a look at the fluidity of the relationships between the two spheres, and how the researchers themselves contribute to this (Fochler, 2016; Lacy et al., 2014; Vallas & Kleinman, 2007).

Twelve semi-structured one-on-one interviews were conducted with machine learning researchers currently working in academia and industry[2]. Questions were asked to elicit responses about researchers' opinions about their choices to leave or remain in academia after their doctorate degrees, and the way in which the two spheres interact in machine learning research. Interviews lasted half an hour on average and were conducted between June and August, 2016. All interviews were audibly recorded and partially transcribed. Ethical approval for this research was granted by the UCL Department of Science and Technology Studies in accordance with its ethical research policies and procedures.

---

[2] "Industry" research is used throughout as synonymous with "corporate", meaning research done at a for-profit institution.

## Interviewees

The interviewees were machine learning researchers currently working in the field in academia or industry, and were based in the United States, Canada, and the United Kingdom. Interviewees were recruited via snowball sampling. All had PhDs, and almost all interviewees were academic natives of the field—that is to say, their doctorates were in computer science with a machine learning focus.

It must be stressed that the interviews conducted are not intended as a representative sample of the field, of researchers in academia or in industry. Naturally there is an issue of self-selection bias at play. Furthermore, snowball sampling was used with interviewees suggesting further participants, again introducing the potential for bias. There are a myriad number of demographic considerations that may be reflected in an individual's response, such as age, academic age, seniority within an organization, gender, nationality, etc., and it is not the goal of this dissertation to assign causational factors to responses.

These limitations are further compounded by the fact that a high-level of anonymity was guaranteed to interviewees in exchange for their participation. The majority of interviewees agreed only to be interviewed on the grounds that, in addition to their names, their place of work be withheld. When referring to specific interviewees, the gender-neutral third-person pronoun "they" will be used, to both protect identity and avoid gender-based inferences. Conclusions cannot be drawn with regards to specific research environments. Instead, all participants offer the views of members currently working in the machine learning community with an interest in its progression.

# Machine Learning in Context

## What is Machine Learning?

Machine learning is a sub-field of research within the field of Artificial Intelligence, part of the discipline of Computer Science. Definitions for all three, along with what they subsume and exclude, have changed over time. The issues inherent with defining "intelligence" so as to circumscribe a set of goals and fields encompassed by the term "artificial intelligence" have been well-documented (Collins, 1987; Fleck, 1982; Fuller & Collier, 2004; Herbert A. Simon, 1991; Williamson, 2010). As Fleck (1982, p. 172) points out, ambiguity in the term AI "is institutionally manifested in the high degree of research area differentiation, with interdisciplinary and multidisciplinary affiliations, and associated multiple funding sources". Here we will not labour over how different definitions of "artificial intelligence" are negotiated, but focus on the research area of machine learning with a brief summary of what it sets out to accomplish[3].

As a subfield, machine learning "has been a central concern in artificial intelligence since the early days when the idea of "self-organizing systems" was popular" (Carbonell, Michalski, & Mitchell, 1983, p. 69). Mitchell (2006) positions it as emerging from the interaction between computer science and statistics. One definition of machine learning from within the field is: "How can we build computer systems that automatically improve with experience, and what are the fundamental laws that govern all learning processes?" (Mitchell, 2006, p. 1). Its applications are extensive, and include image recognition and processing, natural language recognition, digital marketing, self-driving cars, finance and medicine. Some have credited machine learning with the renewed investor interest in AI (Waters, 2015).

---

[3] For a look at different approaches to definitions within the field, see Russell and Norvig (1995).

However, the precise relationship between various methods and techniques of AI and what has been categorized here as "machine learning" is fluid and open to ontological debate[4]. Currently, much of machine learning's focus is based around associated concepts such as neural networks, supervised learning, unsupervised/reinforcement learning, and deep learning (which some call a rebranding of 'neural networks'[5]). One interviewee described the recent popularity of neural networks using a familiar analogy: "It's like discovering a hammer in your toolbox and then hitting everything with it. But most things are not nails" [103].

## A (Brief) History of AI Research

It is necessary to situate the subfield of machine learning within the greater history of artificial intelligence in computer science, so a brief history of the development of the field is offered here. While the course of machine learning's development runs alongside that of AI in general, there are certain deviations specific to the subfield as certain methods become popular and others fall out of favour. In general, however, the short history of the field is one of promise, failure and rehabilitation.

---

[4] In the past some have differentiated between "neural networks" and "AI", describing the former as a competing method to the latter which usually implies "symbolic AI" (e.g., in Guice, 1999). Today, however, they are regarded as a type of AI method.

[5] Accusations of new developments being a "rebranding" of old techniques in AI are not new, e.g., Anderson (1984).

The first recognizable instance of "artificial intelligence" research in its contemporary form was in 1943, although the field itself relies on a variety of concepts pioneered centuries and millennia earlier, including mathematics, philosophy, psychology, computer engineering, operations research and linguistics (Russell & Norvig, 1995; Herbert A. Simon, 1991). The development of computer technology in the 1950s enabled experimentation and the creation of AI tools, which occurred both within industry and academia (Russell & Norvig, 1995). This first era could be characterized by an interest in developing "a general-purpose search mechanism trying to string together elementary reasoning steps to find complete solutions" (Russell & Norvig, 1995, p. 22).

How AI developed as a scientific field in general during this time is naturally of interest, but cannot be delved into too deeply here[6]. Tedre (2015) remarks that computer scientists began to divorce their field from mathematics as it began to develop in the 1950s and that universities had established computer science departments by 1962. Philosopher of AI and professor Hubert L. Dreyfus remarked:

> A mature science progresses by setting forth clear predictions and subjecting them to falsification, then asking why the prediction failed, which assumptions turned out to be sound, and which unjustified, and so learning from mistakes. By such standards AI is only half a science. In the first 20 years—from roughly 1960 to 1980—it had clear goals and forthright predictions as to how and when they would be achieved.

---

[6] For a good summary of how computer science and AI came to be defined in relation to other sciences, see Tedre (2015).

The goal was programming computers to perform in intelligent ways, and the method was to use symbolic knowledge representations. (H. A. Simon et al., 2000, p. 15)

This initial golden period ended by the late 1960s as funding from both British and American government sources began to dry up for AI research, at least partially in response to the failure of these projects to deliver on ambitious goals (Carbonell et al., 1983; Russell & Norvig, 1995; Tedre, 2015). With regards to machine learning, neural networks had run up against computational limitations, and by the end of the 1960s their development in computer science was largely abandoned in favour of symbolic AI (Olazaran, 1993).

The 1970s then saw the development of "expert systems", software used by computers to make decisions or provide advice (such as medical treatments) based on a set of knowledge and facts, and given a set of logical rules applied to these facts to reason out new facts (Russell & Norvig, 1995). Successful expert systems for commercial use were developed in the 1980s and eventually "Nearly every major U.S. corporation had its own AI group and was either using or investigating expert system technology" (Russell & Norvig, 1995, p. 24). Renewed corporate and government interest in the field also came as a reaction to the Japanese "Fifth-Generation" project, which set out to build intelligent machines with ambitious goals like natural language comprehension (Olazaran, 1993; Russell & Norvig, 1995). However, funding for AI research in universities from the US Department of Defense's Defense Advanced Research Projects Agency (DARPA) had begun to decrease, while in the UK the Science Research Council's 'Lighthill Report' criticized the lack of progress made in the field and severely curtailed university research funding (Hendler, 2008).

The mid-1980s also saw the resurgence of interest in neural networks, which had once been replaced by symbolic AI approaches (Hendler, 2008; Lincoln Laboratory, MIT, 1989; Olazaran, 1993; Russell & Norvig, 1995). At the same time, faith in expert systems was waning[7]. Olazaran (1993) attributes the successful re-emergence of neural networks as a prominent research program in the 1980s to a new generation of researchers in various disciplines. Guice writes that "By 1990, virtually everyone active in computer research had heard of 'neural nets'" and that publications in the field had increased ten times since from 1980" (1999, p. 86). However, in corporate environments in the 1990s, AI had become associated with overhype and underachievement (Buck, 1997). Methodology and tools taken from AI had to have the association dropped the association in order to be palatable.

Since the mid 2000's, machine learning has enjoyed a resurgence in popularity with the development of new deep learning/neural network techniques and big data, and with it, renewed corporate interest (Lohr, 2012; Markoff, 2016; "Rise of the machines," 2015; Waters, 2015). One of the most well-known corporate research laboratories in the subfield is DeepMind, which was acquired by Google in 2014 for between USD 400-600 million (Bilton, 2014; "Don't be evil, genius," 2014; Regalado, 2014). Facebook, Twitter, Yahoo, Intel and Dropbox have all purchased AI companies, among others (CB Insights, 2016, "Million-dollar babies," 2016; Kelly, 2014).

---

[7] AI researchers had begun to distance themselves from expert systems as not 'proper' AI see Forsythe (1993); Hendle, (2008); H. A. Simon et al. (2000).

As evidenced by this brief history, the field has undergone several cycles of promise, advancement, and disillusionment. The term "AI Winter"[8] is used to describe periods of little funding, interest, and, seemingly, development in the field (Markoff, 2016; "Million-dollar babies," 2016). As will be discussed later, the historical spectre of the field's cyclic nature is ever-present among researchers. A 2008 letter from the editor of the computer science journal *IEEE Intelligent Systems* warns of the potential for a new AI Winter in the 2010s, despite acknowledging an increased recognition of research and interest in the field, and that corporations had warmed to the term 'AI' again (Hendler, 2008).

# Research Findings

This dissertation focuses on both micro and macro changes occurring in machine learning, affecting individual researchers and the field respectively. These changes are investigated through the recognition that industry and academia are the two leading sources of research and thus a closer look at how they converge is warranted. It is important to note that the distinction between 'industry' and 'academia' is somewhat crude, as neither corporate nor academic laboratories are monolithic. Indeed, the responses from researchers in industry represent a number of different corporate motivations and values dependent on the institution, as will be seen. The following sections examine how expectation dynamics present themselves in the field, how researchers make decisions within their institutional boundaries, how industry and academia influence one another, and what resulting effects have been felt on the research agenda.

---

[8] This is not the only comparison that has been drawn between nuclear technology and AI—for instance the cluster of prominent AI researchers working for Google was compared to the group of scientists assembled for the Manhattan Project ("Million-dollar babies," 2016).

## Expectations of Cyclical Expectations

Worries that industrial research labs draw a disproportionate or untenable number of researchers away from academia have been the subject of media attention in AI since the 1980s (A.E.S., 2016; Bronner, 1998; Gibney, 2016; Hart, 1982; "Million-dollar babies," 2016; Regalado, 2014; Weiner, 2000; Wilson, 1999)[9]. However, almost all the interviewees indicated they were not concerned about it when asked, viewing industry instead as a positive by providing more jobs for graduates. Half of the interviewees mentioned that there was an insufficient number of faculty positions available in any event. One remarked that academia was set up like "a Ponzi scheme" [104]; Stephan and Levin (2002) noted that this pyramid system works so long as funding grows correspondingly to provide jobs for the increasing number of graduate students. While the historical narrative has been that computer science departments in the 1980s were unable to hire enough staff to keep up with student demand, no data suggests that this is currently the case (Roberts, 2011). A few interviewees stated that nearly all their PhD colleagues had gone to work in industry, with some concentration at certain companies. Said one industry researcher:

I think that they [students] can feel frustrated if too many of their peers go

immediately to [industry] labs… I think that sometimes labs can easily get sort of a

---

[9] Evidence of concern over the US military attracting researchers with its own well-paying research jobs was not as apparent, but still present (Joyce, 1984).

bad reputation if they're not publishing quickly enough, if they're not being open

enough. It's a very quickly changing world in that respect. [111]

Google DeepMind is a prominent example of a lab which has hired an increasing number of

researchers since its acquisition. *Nature* estimates that it currently employs at least 144

researchers, 65% of whom researchers were hired directly from academia (Gibney, 2016).

The majority of interviewees expressed a belief that this hiring pattern was cyclical,

and that as machine learning became less attractive to companies, many would return to

academia[10]. "I think it's a normal cycle and I'm sure in a few years things are going to go on

the other way again that the economy is not going to stay like this" [109]. Only one

academic researcher expressed a concern that there wouldn't be sufficient supervisors to

handle the influx of students at universities. Another academic mentioned that recent hiring

at the university had been successful, if not a little more difficult than before, "but every

time I mention this to someone they almost always say it's always been hard to hire, so it's

hard for me to know if that has really shifted" [105]. As one industry researcher put it:

> … it's happened before, the field of computer chip manufacturing was once
>
> something that was an academic field, and now it's not. AI is becoming that, that's all.
>
> I'm not worried that generations of electrical engineers have been going into
>
> industry. [107]

Studies in expectations dynamics have produced different frameworks for understanding these cycles. 'Expectations' here refers to present representations of future capabilities of a technology, and 'hype' is a wave of "high rising expectations" (van Lente, Spitters, & Peine, 2013, p. 1615). The 'hype-disappointment' cycle is used to describe how an emerging technology's capabilities are initially hyped to procure funding and support, followed by the concomitant disappointment and withdrawal of support when expectations are not met, and the eventual recovery over time as the technology advances and becomes adopted (Kirkels, 2016). Case studies have revealed a more complex progression, with disappointment at one level not necessarily translating into a withdrawal of support at others (van Lente et al., 2013). Nevertheless, while this model may be critiqued on the grounds of being overly deterministic (Borup et al., 2006), the description of machine learning's history as a series of successive hype-disappointment cycles appears to be an accepted *narrative* amongst different actor groups. This is evidenced by interviewees' common referral to cyclicality in the field, as well as the historical and present media coverage and funding decisions presented throughout this dissertation. Remarking on the current environment, an older academic researcher said:

> … if you've been through the previous iteration you've seen the rise and then the crash and how uncomfortable and difficult it was on the crash side, it's hard to see the rise right now and not ask, are there things that I could be doing more of that will soften the landing on the other side of this thing. [109]

---

[10] For a historical look written immediately after the dot-com crash about faculty migration, see Stankovic and Aspray (2003).

Furthermore, expectations can manifest differently among actor groups (such as researchers and funders) and also at different levels, including micro (research groups), meso (the field), and macro (society) (Borup et al., 2006; van Lente et al., 2013). A survey of 700 neural network researchers found differences in the factors which influenced their decision to work in the field between those who had worked in it prior to a period of hype and those who entered it during (Rappa & Debackcre, 1990). The latter group was more compelled by "recent successes", "positive opinions", the "availability of funding," a "lack of other topics" and "financial rewards" than the former, which were more likely to rate "intellectual compellingness" as more important.

As this sample is not representative and only one interviewee expressed concern, it is impossible to say if this attitude currently differs among industry and academic actors, as some studies of other technologies have found (Kirkels, 2016). Researchers' public denunciation of hype—particularly as propagated by the media and industry investors—can be traced back to at least the 1980s (Anderson, 1984; Guice, 1999). This aversion to hype has been attributed to the fear that heightened expectations among funders (including government, military and industry) invariably precipitates disappointment, and subsequently AI Winters (Anderson, 1984). In its study of neural networks in 1987-1988, DARPA interviewed hundreds of researchers in the field and produced a 600-page report (Lincoln Laboratory, MIT, 1989) which addressed the hype surrounding the field in cautionary tones (Guice, 1999). DARPA intentionally framed the report as a dispassionate look at the field and a presentation of objective reasons as to why it should receive military funding.

## Acts of Agency

When discussing why they chose to work in either industry or academia, freedom and choice were mentioned as the primary factors in the decision by researchers in both spheres. The advantages offered by either were things that enabled expressions of agency, and the disadvantages acknowledged the institutional boundaries limiting it. Often, these enablers and constraints took the form of resources, including financial (personal compensation as well as research funding), infrastructure (engineering and computational), informational (data), and human (colleagues, graduate students). This decision was also sometimes accompanied by a desire to simply 'try something different'.

One researcher, who had been working in industry since their PhD, was returning to academia after a change in corporate direction to a focus on external clients had left them unhappy. This corporate shift resulted in changes to the types of research being pursued by the lab, which had become more short-term in nature. About the expected benefits of academia, they remarked:

> Another downside, or upside, or both, is I may have more freedom to pursue the research I believe in, but at the same time I have the responsibility to get funding for it. At [company] there were other people who had to worry about funding and I didn't really face as much funding pressures there as I will face here [university]. [112]

The recognition of this trade-off and the similarities of both institutions has been noted by Shapin (2008) as historically occurring. Furthermore the response echoes Holloway's examination of the ways expressions of agency are presented within the boundaries established by commercialization: "To assume that all scientists unwittingly accept commercialization is to obscure the diversity of their experiences, including the various expressions of their agency" (2015, p. 745). At the same time, for Holloway the neoliberal prioritization to commercialize established the boundaries of agency by promoting and encouraging those options which supported this goal. In looking at what motivates scientists engaged in commercial research, Lam stated: "Scientists' engagement in commercial activities will need to be interpreted within this shifting institutional context in which individual action often reflects the contradiction experienced rather than necessarily signalling unequivocal acceptance of a particular set of norms or values" (2011, p. 1356).

With regards to current employment mobility, there was no indication that interviewees saw any obstacles preventing movement from one sphere to the other. Two of them were on loan to industry from academia and another felt that dual affiliations were generally a positive, as well as a potential guard against academic brain drain:

> I think it's really common right now, a lot of these big names that are leaving machine learning aren't really leaving academia, they're taking leaves of absence and are still a professor somewhere. But I mean it's an exciting time, one reason why people leave and it's similar for me, is that the resources are different in industry… [108]

Some interviewees felt academia was burdened with too many administrative duties, while others expressed the same feeling with regards to their industry jobs. Dismay at the administrative nature of academia has also been noted elsewhere:

Graduate students in the sciences have expressed disappointment upon discovering the high levels of entrepreneurialism required to keep a university laboratory afloat, and many have decided instead to pursue careers in high-technology industry, where ironically they believed they would have more autonomy and flexibility, experience less pressure than in academia. (Moore et al., 2011, p. 513)

This awareness about the balancing act between resource access and control over the research agenda was expressed by all interviewees, but not necessarily in expected ways:

The corporate interests are there and the pressures are there but honestly I find, in terms of the direction that they exert on my research, I find them less pronounced in industry than I did in academia because in academia you're always under pressure to produce the next paper. And so there is a huge pressure to get results that will secure you a publication and then have enough publications to write your next form that you have to fill in for funding. [104]

Freedom from funding concerns was mentioned as one of the main benefits of working in industry. The pressures on academics to obtaining funding was generally understood and acknowledged by researchers in both spheres and has been noted elsewhere, along with the desire to avoid the issues of procuring funding as a motivating factor among researchers to leave academia (Vallas & Kleinman, 2007). For other respondents, however, industry's explicit set of constraints was not necessarily a positive.

Researchers in both academia and industry expressed an aversion to military funding, and one who had previously worked at a non-profit that received military funding said it factored into their decision to leave for a job in industry. Historically the military has been a large source of funding for computer science and AI research, and was often non-directive driven (Fleck, 1982; Olazaran, 1993; H. A. Simon et al., 2000; Smit, 1995)[11]. Interestingly, other projects which began life as military research programs, such as the voice command system Siri, were spun-off and acquired by corporations (Apple in this case) (Lohr, 2012). This raises the question of when military funding fell out of favour among researchers. One possible explanation is that researchers in industry are no longer reliant on it (Dougherty, 2016). An industry researcher said it seemed "secondary" as a funding source because "the level of corporate money is so huge that I think it dwarfs even what the military are going to put into something like this" [104]. Perhaps researchers simply feel more comfortable voicing their discomfort with it because there are alternatives. This aversion may cause difficulties for academics in obtaining funding.

One of the ways corporations shield their researchers from short-term profit-oriented directives is through the establishment of "skunk works" (Gwynne, 1997). These refer to research groups which pursue their own research agenda largely (or totally) without the oversight of managers and corporate bureaucracy, and thus free from the usual constraints placed on industry research and development. The purported benefit is a chance for longer-term oriented innovation which would not usually be considered tenable, and a more flexible approach to research (Gwynne, 1997). However, these organizations may foster a false sense of security and may find themselves in a precarious position, subject to the whims of the market and/or management. One of the interviewees had been working for an industry research lab but returned to academia when they were laid off along with the rest of their group, which was comprised of machine learning "superstars". Initially operating as a sort of isolated skunk works, the lab had become no longer immune to the company's financial considerations. Reflecting on the trade-offs between industry and academia, the researcher said:

> I had access to tremendous resources because the company was very very rich but anything I wanted I had to ask for it... I would go my boss and I'd say 'would it be okay if I got a computer' and he would generally say 'sure', but the dynamic is different as a professor—you're responsible for really running your own show and so I have a lot less resources but it's really up to me to decide how to allocate them. [109]

---

[11] For instance, Fleck (1982) puts 75% of funding for AI research between 1964-1974 in the US as coming from military sources.

Industry labs may still enjoy this protected status so long as their parent companies generate enough profit from their other ventures[12]. But the ability to do so is situational and dynamic. Nelson (2002) states that corporations whose products depend on a narrower range of technology will be more inclined to fund research which can quickly be applied, whereas those with a broad scope of products dependent on a broader range of technologies may be able to fund more basic research.

---

[12] For instance Alphabet, parent company of Google, funds a skunk works division called "Other Bets" which boasts increasing operating losses while other divisions post increasing profits (Barr, 2016).

"On one level, scientific work is an oscillation between intense communal interaction and solitude" (Henke & Gieryn, 2008, p. 361), and from the interviews there is evidence that researchers negotiate within the environments of their research space how much of both they encounter, and it seems that certain types of labs, along with level of seniority within a corporation, grant researchers more opportunities to work in the mode more suited to them. A few researchers hypothesized that the hierarchy of larger labs, such as Google DeepMind, might prohibit some autonomy from setting the research agenda, and thus they were more attracted to working for smaller companies like start-ups. Others felt academia had afforded them more time to just think about problems and less pressure to come up with successful research quickly. One mentioned missing the academic environment, as it allowed for interactions with others from a wider array of disciplines.

On the other hand, an academic mentioned that access to a team with specialities, such as research engineers, provided by working in industry would allow them to spend more time on reading and research and less on implementation. Some industrial researchers brought up the relatively solitary nature of their PhDs in comparison to the lab environments they now found themselves in, where they had access to researchers and engineers who were specialized and were expected to work more within a team.

Some researchers enjoyed teaching, while others enjoyed the opportunity to mentor graduate students provided by academia. Slaughter et al. (2002) wrote about the graduate student as a resource and how their role as currency between academia and industry. Two interviewees, both in industry, affirmed the graduate student role as capital in this trade, with one ruminating on its positives and the other viewing it as more exploitative:

I'm very fortunate that I have access to good interns here so every year I have between 3-4 interns from top universities so in that sense, it's better, because without being a top university you have access to that calibre of students, but of course I have them for a very short time… [109]

Universities sort of vie with one another to attract the attention of companies so as to get enough funding for their students or they become almost like suppliers of grad students for companies and I think none of those things are as all good for universities long-term. [104]

Access to informational and infrastructure resources were acknowledged as an important consideration for what types of research could be conducted. Fleck (1982) viewed "adequate computing facilities" as part of the paradigmatic structure of AI (in the Kuhnian sense of paradigm), and made its funding a critical component of the field's progress. Data plays a large role in the success of current deep learning techniques, and companies such as Google and Facebook have a seemingly endless supply ("Facebook, Imperial ambitions," 2016, "Rise of the machines," 2015)[13]. An interviewee said:

…there are things I think at the moment honestly you can only do in industry for machine learning. Particularly if you want to use very large data sets then industry has access to very large data sets and very large computing infrastructure on a scale that you just can't build in academia. [101]

---

[13] The importance of data as "capital" has also been observed in pharmaceutical fields (Fochler, 2016).

Others in industry downplayed resource access advantage though, focusing on the fact that for findings to be meaningful they had to be proven against other research benchmarks on publicly available datasets.

## Negotiating Converging Norms

For Vallas and Kleinman (2007), high-technology industries have adopted many of the norms and practices traditionally associated with academia, and vice-versa. This has been termed "asymmetrical convergence" to indicate that, while the two spheres may be increasingly mirroring each other, it is ultimately in service to corporate interests, a neoliberal conclusion. If the spheres of academia and industry have become semi-permeable in the eyes of researchers, how does their mobility across them affect the culture of each? What do researchers bring with them, and what are they forced to leave behind? This section will examine the ways in which tensions between the two are navigated with regards to forms of academic recognition, machine learning conferences, researchers' sense of community and the effects of industry research on the direction of the field.

### Academic Recognition

Vallas and Kleinman viewed the adoption of academic norms by industry with suspicion: "In private sector science, the free flow of ideas—a deeply held academic ideal—is promoted, but within constraints and always in the service of profit" (2007, p.

289). The interviewees' responses suggest that this type of business strategy can be effective. Publishing in academic venues (including journals and conference papers) was still important to all interviewees working in industry, although some, particularly those at start-ups, recognized it was not necessarily a feasible focus. It was understood that companies had an interest in allowing their researchers to publish to attract other researchers and maintain an image of being on the cutting edge, and an academic researcher mentioned they thought it was seen as "prestigious". In the literature publications have been identified as a form of capital used to attract resources which would help a company's longer term goals (Fochler, 2016). In fact, companies such as Apple have been criticized by researchers both in industry and academia for not publishing, which is often taken as indication that their machine learning research program has not developed very far (Levy, 2016). Lam (2011) found that "traditional" academic capital in the form of recognition through publications was still an important motivator for a large percentage of scientists engaged in entrepreneurial activities.

One researcher in industry mentioned that, while working on new features for applications was the company's first criterion, publishing was encouraged and researchers received financial incentive to do so. A few in industry mentioned that the corporate setting could slow down the publishing process but not prohibit it. One saw the ability to publish as the "prize" for their start-up being successful [108]. Another industry researcher expressed an awareness of internal tensions between their commitment to science and working in a corporate environment:

I don't want to make this seem like a moralizing fact but I did kind of buy into the thing when going into science that a scientific discovery can be one of the most important things that ever happens in the world… I think I have kind of a phobia of slipping into being someone who cares about companies. I just don't give a shit about [company] or anything. I hope that they stay around so that I have a job, and it's a great place to work at… but I really hope that I never become someone who really cares about user experiences or making sure that there is a higher recall precision. [107]

For other researchers, academia still carried with it a certain cache, reminiscent of the valorized scientists of Merton's scientific norms (Shapin, 2008):

The only things I don't like about industry are I have a certain sense of guilt about being part of a large corporation and working for the man and kind of somehow I feel like academia is a more noble pursuit but this is quite a vague feeling. [104]

I'm just in love with being a scientist. I think that's my favourite part of the job is trying to discover knowledge and trying to be really really careful, trying to set up really nice experiments, I don't really see myself as a computer scientist or engineer, I really do see myself as a scientist… [108]

**Conferences**

Conferences have long played an important role in machine learning, with the two largest and best-known in the subfield being Neural Information Processing Systems (NIPS) and International Conference on Machine Learning (ICML). *The Economist* called NIPS "the Davos of AI" and stated that from 2010 to 2015 attendance tripled to 3,800 ("Million-dollar babies," 2016). An increase in attendance and industry presence was unanimously given as one of the largest changes to conferences in recent years, and they had become a site for recruiting among companies. This was seen as both a positive and negative among interviewees:

> They [industry] make a lot more papers and the quality of the papers is very high, in some respects there's certain kinds of research that basically is way better in industry than it could possibly be done in academia. They try to attract PhD level people so they have in some ways a lot more freedom for these people to pursue research although the research is often very much encouraged to be aligned with the company. [101]

If academia was once seen as the place where machine learning research needed to come from in order to be recognized as "legitimate" (Henke & Gieryn, 2008), several responses suggested that conferences were evidence this legitimacy had been extended to industry:

> You have a far larger number of people of course from industry presenting at those conferences. It's kind of surprising that researchers at Microsoft or Google are going to the lectern and showing slides on abstract things because only a few years ago it seemed like the researchers of those people were second tier at some extent to the academics, but now it's maybe shifted a little bit. [107]

…all I've heard is basically positive responses from current PhD students because it's not like we're trying to distract, we're doing core research as well. […] I think it would be worse if the industry was coming in and being distracting and saying 'work on this application, work on search, or work on this', but I don't see that happening. [111].

Researchers also expressed concern that conferences weren't as much fun anymore, and that it was more difficult to have the type of one-on-one interaction with other attendees, which had prior been one of their main benefits. The increased visibility of industry research appears to be salient:

The big academic conferences now look almost like industry fairs with all the stands for the big companies all of them recruiting, a lot of the papers you read are now published by companies. The companies are now dictating the tempo definitely. [104]

Some have suggested that increased industry participation may force computer science research to grow out of its conference centric state, along with the field maturing (see Fortnow, 2009).

There is a familiar theme to be found among the changing status of conferences. From a 1997 *Computerworld* article: "The annual American Association for artificial intelligence (AAAI) was a raucous, lavish affair during the 1980s. The most recent AAAI meeting, which was held in Portland, Ore., last August, was subdued by contrast" (Buck, 1997, p. 80). (Rappa & Debackere, 1989) spoke of the need for informal spaces where AI researchers could communicate easily with one another and how the growing size of the research community hindered this.

## Community

Interviewees affirmed the importance of having a sense of being part of the machine learning community, and engaging with its members regardless of their affiliation. The idea that scientists in the same research area comprise an "invisible college" where they spread knowledge through their interactions with their peers is not new (Powell, Grodal, Fagerberg, & Mowery, 2006). Polanyi (1969) asserted that scientists exhibiting agency in research were also acting in co-operation within their associated communities, and the recognition of that community displayed here is important to consider alongside the previously pros and cons of each sphere. A researcher mentioned that, due to the growth in corporate machine learning research and its ability to attract top researchers, going to work in industry no longer meant leaving the community behind. One industry researcher felt that their company had to "be conscious of paying back to the community" and aware of its influence [111]. For another it was part of why they chose to work for a specific company:

> …that's another reason why I chose to work here [company] because they are sort of focused on not just working on kind of interesting things but also allowing me to continue to let other people know what I'm doing, and so maintain my participation in the academic community – personally, I enjoy that. [106]

This could also be taken as further evidence of convergence between the two spheres, and Vallas and Kleinman characterize industry's increasing "concern for collegiality" as a means to recruit and motivate researchers rather than as an end (2007, p. 289). It is possible that certain organizations do care about their researchers maintaining this sense of community for a variety of other reasons, while their competitors do so because they realize it is a necessary condition to attract talent.

It may be that the tightly knit sense community in machine learning, along with its set of norms, may undergo a change due not only to increased industrial interest but an increase in the number of its practitioners, which may result from a variety of causes such as decades of neoliberal governmental policy promoting "STEM" careers along with the level of compensation currently offered by the field. Currently, it may be the machine learning field affiliation, rather than an academic/industry divide, which unites researchers under shared norms and history.

**Effects of Industry Research on the Field**

Given the presence of industry research in the field and its acceptance among researchers as a legitimate source of 'academic' research, it is only natural to wonder to what extent it has altered the research agenda in the academic sphere. When discussed with the interviewees, there was some disagreement on how corporate influence presented itself and was felt. One academic addressed how methods used and publicized by corporations were affecting the direction of publications in the field:

> I see more things from reviewers like 'you didn't run experiments on Atari, we should reject this paper'. Which is something that I would never agree with, you don't need to have experiments on Atari. That might not necessarily be a DeepMind person [the reviewer] in fact, it's just that the influence that they have brought to the field is just to make people think that in order to have meaningful research you have to have these big results on domains like Atari. [105]

An industry researcher offered this perspective:

> Whether or not people are reading [industry] papers and being like 'oh I should work on that too', or they've been introduced to a similar system of thinking and they're starting to ask the same questions, I can't distinguish entirely and maybe it's silly to. [107]

Still, with regards to the methodologies and experiment domains that had received attention in recent years, such as those popularized by Google Deepmind,[14] many expressed the belief that these things were cyclical in the field, and eventually a new "fad" would become popular. At the same time, it was recognized as potentially "professionally damaging" to completely ignore what was currently en vogue.

All interviewees agreed that academic research remained and important source of knowledge and findings in the field. One felt that quality research was more dependent on the individual than their organization. It is clear that personal interest played a role in the type of research a researcher would pursue, within the abilities of the institution to accommodate them. Many interviewees mentioned that the type of research they were pursuing had not fundamentally changed when they made the move from academia to industry:

> Really, it's remained on the same path. I mean I've been influenced of course by the researchers around me, but I always was, before in academia as well, but mostly I'm still sort of continuing and pursuing the same kinds of things that I was before. [104]

---

[14] Specifically deep learning/neural networks applied to domains such as Atari and Go (Mnih et al., 2015; Silver et al., 2016).

This same industry researcher predicted that the field was going to become more "application-centric". However, the interviewee also later expressed excitement at the fact that AI was moving into applications to catch up with the theory, and that in the future renewed focus on theory could be necessary, again taking a cyclical view on the field. Similarly, one researcher said that the focus now seemed to be on "system building" and that was a sign that the field was maturing past figuring out the fundamental building blocks—it was now about combining them [108]. Moore et al. writes:

> scientific fields tended to become structured by a tension between a producer pole that emphasized independence and traditional scholarship, and a practical pole that emphasized new collaborations and industrial applications. The dominant pole varied across disciplines (Albert 2003). (2011, p. 513)

> One industrial researcher said:

> …we may be hearing more about these paths [industrial research] right now because they're either having more success from a practical perspective or maybe they're having more success in the press, but ultimately I think all the research that people were doing before is still very valid, there's actually very few research projects that were being done in academia now whose course has been altered by the added interest in machine learning in industry. [103]

A third of interviewees made the theoretical/applied divide with regards to the type of research that was, or could be done, in academia and industry respectively. Researchers at start-ups acknowledged that their research goals were much more application-focused than they had been in academia. Industrial research was seen as 'getting things to work' by academic researchers, although this does not necessarily mean in an "applied" way, whereas academia was about understanding smaller problems, such as the specifics about how an algorithm worked:

…they're [industry] not nearly as concerned with understanding things, it is that kind of mentality that you have to produce stuff and stuff has to work and that's great but at some point you also have to understand why it works. [101]

This account was disputed by a researcher in industry, however, who enjoyed undertaking the analysis of why something had worked. It may be that particular corporate lab environments are more supportive of post-mortems than others.

Some industry researchers felt that academic research should focus on "the things that business isn't interested in it… the sort of big ideas for which there is no obvious financial motive" [104] rather than compete in the same domains as corporate research. But this was also accompanied by the awareness that obtaining funding for such research in academia was a problem. Both industrial and academia mentioned the "incremental" nature of the research in academia. The narrative of "applied" vs "theoretical" research as the delineator between industrial and academic research respectively appears to be slightly more complex, with perhaps "incremental" being useful as a distinction.

# Conclusions

Rather than setting out to collect evidence to support or disprove a set of hypotheses about the changes taking place in machine learning, this dissertation assembles a collection of viewpoints from the researchers themselves. The responses are both deeply individual and potentially indicative of wider trends.

Researchers discussed the ways in which they navigated the limits of their institutional spheres with respect to their own agency. The opportunity to conduct research they were personally interested in was emphasized as important, with the recognition that they were operating under certain pragmatic restraints. The findings here also offer support for what has been observed elsewhere, that "with important cross-firm variation, the organizational culture and practices we unearthed [in corporations] drew freely from academic norms and conventions" (Vallas & Kleinman, 2007, p. 294).

Furthermore, the cross-pollination of academia and industry could be seen in the ability of researchers to navigate between the two, along with some indication of convergence between them. There is much discussion and debate as to how the increasing presence of industry at conferences and in publications is directing the research agenda, as well as how much of this is a product of a historically-occurring cycle. The skepticism of hype and seemingly stoic resignation that the current level of interest in the field will not last forever provides an interesting perspective on the dynamics of expectations.

This dissertation leaves much unexplored, and, given the relatively small amount of contemporary literature written about machine learning researchers, there are many directions for future research to take.

## Future Research

This dissertation conspicuously avoids issues related to gender, class and race in its discussion of both the field and identification of interviewees. Saxenian (1994) argued that the tightness of the community of engineers in Silicon Valley was partly due to their homogenous (young, white, male) identity. For a discussion of how gender in corporate machine learning research environments might be approached, Slaughter (2001) would be instructive.

Furthermore, this dissertation only touches lightly on the implications of the science/technology and basic/applied research divide, which is sometimes superficially assumed to characterize the difference in research done in academia and industry respectively. This has been challenged and discussed elsewhere in STS (e.g., Berkhout, van der Duin, Hartmann, & Ortt, 2007) and applying it to the interviewee's responses would be interesting.

AI has long been a source for scrutiny about the safety of autonomous robots, and more recently questions of the ethical implications of the application of machine learning algorithms in domains such as policing have been raised (e.g., Knight, 2015). Questions of ethics, responsibility and how the field has attempted to govern itself through various organizations, both academic, corporate, public and private were put to the interviewees but are not covered in this dissertation. Their responses, however, are indicative of the various ways in which researchers exhibit reflexivity, and defer or accept issues of responsibility (e.g., as discussed in Stilgoe, Owen, & Macnaghten, 2013).